# Spontaneous Symmetry Breaking: The Case of Crazy Clock and Beyond

Maja C. Pagnacco[1*,] Jelena P. Maksimović[2], Marko Daković[2], Bojana Bokic[3], Sébastien R. Mouchet[4a,5], Thierry Verbiest[6], Yves Caudano[4a,4b] and Branko Kolarić[3,7*]

[1]University of Belgrade, Institute of Chemistry, Technology and Metallurgy, Center for Catalysis and Chemical Engineering, Njegoševa 12, Beograd

[2] University of Belgrade, Faculty of Physical Chemistry, Studentski Trg 12, Belgrade, Serbia

[3]Center for Photonics, Institute of Physics, University of Belgrade, Pregrevica 118, 11080 Belgrade, Serbia

[4a]Department of Physics & Namur Institute of Structured Matter (NISM), University of Namur, Rue de Bruxelles 61, 5000 Namur, Belgium; [4b] also Namur Institute for Complex Systems (naXys)

[5]School of Physics, University of Exeter, Stocker Road, Exeter EX4 4QL, UK

[6]Molecular Imaging and Photonics, Department of Chemistry, KU Leuven, Heverlee, Belgium

[7]Micro- and Nanophotonic Materials Group, University of Mons, Place du Parc 20, 7000 Mons, Belgium

*Corresponding authors e-mail: maja.pagnacco@nanosys.ihtm.bg.ac.rs

branko.kolaric@umons.ac.be

**Abstract:** In this account, we describe the crazy-clock phenomenon involving the state I (low iodide and iodine concentration) to state II (high iodide and iodine concentration with new iodine phase) transition after a Briggs-Rauscher (BR) oscillatory process. While the BR crazy-clock phenomenon is known, it is the first time that crazy-clock behavior is linked and explained with the symmetry-breaking phenomenon, highlighting the entire process in a novel way. The presented phenomenon has been thoroughly investigated by running more than 60 experiments, and evaluated by using statistical cluster K-means analysis. The mixing rate, as well as the magnetic bar shape and dimensions, have a strong influence on the transition appearance. Although the transition for both mixing and no-mixing conditions are taking place completely randomly, by using statistical cluster analysis we obtain different numbers of clusters (showing the time-domains where the transition is more likely to occur). In the case of stirring, clusters are more compact and separated, revealed new hidden details regarding the chemical dynamics of nonlinear processes. The significance of the presented results is beyond oscillatory reaction kinetics since the described example belongs to the small class of chemical systems that shows intrinsic randomness in their response and it might be considered as a real example of a classical liquid random number generator.

**1. Introduction**

The presence of symmetry around us inspired many scientists to search for beauty, harmony, order, and regularity in Nature and Her fundamental laws [1,2]. Additionally, phase transitions with and without spontaneously broken symmetries are widespread concepts through different areas of physics and physical chemistry. The applications of spontaneously broken symmetries cover a wide range of condensed science topics, such as superconductivity, super-fluidity, Bose-Einstein condensation, nucleation physics, self-assembly processes, morphogenesis, and chemical kinetics. In this account, we describe spontaneous symmetry breaking in the case of the nonlinear Briggs-Rauscher reaction. We highlight the importance of symmetry breaking, in non-equilibrium and pattern formation processes, which is of vital meaning to the understanding of the morphogenesis process and for applications in broad areas of biomimetics and nanoscience.

The Briggs-Rauscher (BR) [3] reaction is a hybrid oscillating reaction formed by coupling two chemical oscillators, Bray-Liebhafsky [4,5] and Belousov-Zhabotinskii [6]. Since its discovery in 1973, the Briggs-Rauscher oscillating reaction has been one of the most investigated oscillatory systems. It is probably due to its simplicity and exciting colour alternation caused by changes in reaction kinetics (when starch is used as an indicator) [7].

BR reaction typically occurs within mixtures of $H_2O_2$, $H_2SO_4$, and $KIO_3$. Additionally, Mn(II) ions are added as a metal catalyst and malonic acid ($H_2MA$) as an organic substrate. Substitutions of chemicals are possible; different acids, organic substrates, and ions, such as Ce(III) instead of Mn(II) catalyst, can be used to generate BR oscillations [7-10]. However, the oscillatory behavior is not the only one that attracted the attention of non-linear scientists in the Briggs-Rauscher reaction [11-15].

Indeed, as described elsewhere [16], after the well-controlled initial oscillatory behavior, the reaction becomes chaotic. Depending on the initial conditions, particularly on the ratio $[H_2MA]_0/[IO_3^-]_0$ [16,17], the reaction exhibits a sudden and unpredictable phase transition. This transition, from state I (low concentration of iodide and iodine) to state II (high concentration of iodide and iodine), happens randomly in practice, as the time spent by the system in the state I is irreproducible (see Figure 1). The transition is characterized by a "sharp and sudden" increase of iodine and iodide concentration, followed by the formation of solid iodine. The observed stochastic feature, called a crazy clock (due to the unpredictable time needed to provoke the transition), is linked to imperfect mixing that affects convection and diffusion dynamics. The imperfect mixing results in extremely complicated phenomena, which occur on multiple length and time scales [18]. Possible kinetical consequences are the appearance of bifurcation, chaos, intermittent behavior, and symmetry-breaking [18,19]. Therefore, this paper further studies the mixing effects in connection to the crazy-clock phenomenon in the Briggs-Rauscher oscillatory reaction. It compares and processes statistically more than #60 experiments obtained under identical initial concentrations of all reactants. Although the BR crazy-clock phenomenon was previously detected [16], this behavior is linked for the first time to symmetry-breaking, highlighting the entire process in a novel way. Furthermore, the investigated crazy clock exhibits a truly random behavior that might be considered as an example of a classical, liquid random number generator. Additionally, the investigated system also belongs to the particular class of classical systems that shows intrinsic randomness in their response (as also observed in colloid particles placed on an oscillating surface [20,21].

## 2. Materials and Methods

*2.1 Briggs-Rauscher experimental setup*

Since the time of the transition between state I and II is unpredictable, great attention must be paid to the experimental procedure. Only analytically graded reagents without further purification were used for preparing the solutions. Malonic acid was obtained from Acrōs Organics (Belgium), manganese sulphate from Fluka (Switzerland), perchloric acid, potassium iodate, and hydrogen peroxide from Merck (Germany). The solutions were prepared in deionized water with specific resistance 18 MΩ/cm (Milli-Q, Millipore, Bedford, MA, USA).

All experiments were done in a container not protected from light. Reactions were monitored electrochemically (unless specified in the text). An $I^-$ ion-sensitive electrode (Metrohm 6.0502.160) was used as the working electrode and an Ag/AgCl electrode (Metrohm 6.0726.100), as the reference. During the experiments, the temperature of the reaction container was regulated by a thermostat (Julabo ED, Germany) and maintained constant at 37 °C. The reaction mixture was stirred by magnetic stirrer (Ingenieurbüro, M. Zipperrer GmbH, Cat-ECM5, Staufen, Denmark).

Four independent series of measurements were carried (they differed in stirring bar size and shape, as well as mixing rate) with the identical solution composition $[H_2MA]_0 = 0.0789$ mol/dm$^3$, $[MnSO_4]_0 = 0.00752$ mol/dm$^3$, $[HClO_4]_0 = 0.03$ mol/dm$^3$, $[KIO_3]_0 = 0.0752$ mol/dm$^3$ and $[H_2O_2]_0 = 1.176$ mol/dm$^3$ in 25 ml volume:

1) without mixing (number of conducted experiments #30)

2) with mixing 100 rpm using cylindrical stirring bar 10 mm length, 4 mm diameter (BRAND magnetic stirring bar, PTFE-coated cylindrical), (number of conducted experiments #30)

3) with mixing 100 rpm using cylindrical stirring bar 20 mm length, 6 mm diameter (BRAND magnetic stirring bar, PTFE-coated cylindrical) (in triplicate)

4) with mixing 100 rpm using triangular stirring bar 12 mm length, 6 mm diameter (BRAND magnetic stirring bar, PTFE-coated triangular) (in triplicate)

*2.2. Statistical processing and Cluster analysis*

The obtained experimental results were analyzed in the open-source statistic software "R" using "hclust" algorithm for the hierarchical cluster analysis (HCA) [22,23].

**3. Results and Discussion**

*3.1. Effects of the stirring bar shape and dimensions on the state I→state II transition*

The BR oscillatory period is strongly reproducible, while the transition from state I to state II occurred practically randomly (Figure 1), as previously reported by our research group [16].

It is imperative to emphasize that in our measurements (Figure 1), unlike in other crazy-clock reactions found in the litterature [18,19,24], large time fluctuations (the order of magnitude could be more than two hours) occur after a highly reproducible oscillatory period. Two independent measurements and consequently obtained BR oscillograms exhibit identical trends in oscillation amplitude and time between to neighboring oscillation maxima $\tau_{n-(n-1)} = t_n - t_{n-1}$, as it can be observed in Figures 1, 2a and 2b. Conversion from higher to lower potential of iodine electrode (or from low to high iodide concentration) marks the transition from state I to state II (state I→ state II). Furthermore, the choice of the working electrode affects only the transition shape. However, the transition itself is very noticeable due to the intense color change of the system from colorless to yellow accompanied by solid iodine formation. This allows monitoring the state I→ state II transition with the naked eye.

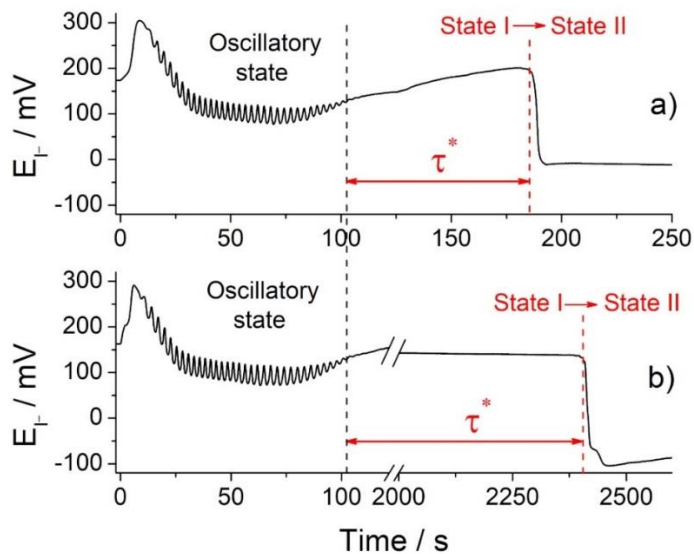

**Figure 1.** Two independent measurements a) and b) of iodide potential vs. time obtained for BR reaction under experimental conditions: $[H_2MA]_0 = 0.0789$ mol/dm$^3$, $[MnSO_4]_0 = 0.00752$ mol/dm$^3$, $[HClO_4]_0 = 0.03$ mol/dm$^3$, $[KIO_3]_0 = 0.0752$ mol/dm$^3$, $[H_2O_2]_0 = 1.176$ mol/dm$^3$, T = 37.0°C. The experiments were performed without stirring and without protection from light. $\tau^*$ denotes the time from the end of the oscillatory mode to the occurrence of state I→state II transition.

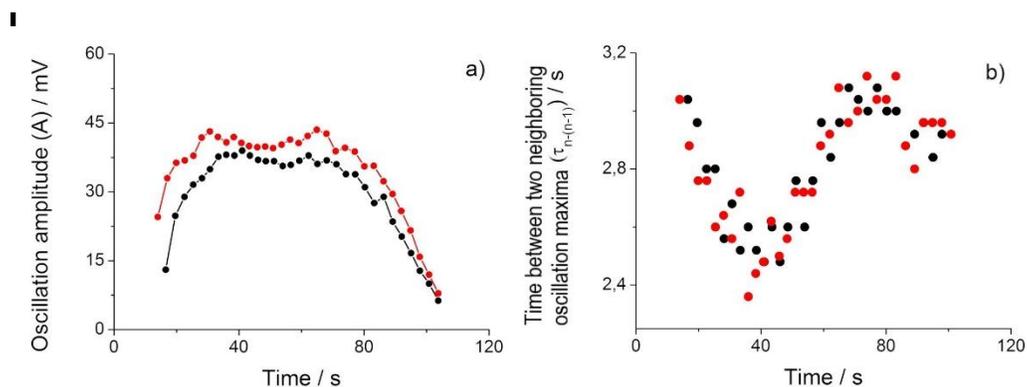

**Figure 2.** Briggs-Rauscher oscillation amplitude a) and time between two neighboring oscillation maxima $\tau_{n-(n-1)} = t_n - t_{n-1}$ b) in the two independent measurements presented at Figure 1. The resulting BR oscillograms have the same number of oscillations ($N_{osc} = 33$) and identical oscillation period, however the time of state I→ state II transition differs more than 10 times (as shown at Figure 1).

The cause of this unexpected transition is still unknown. Previous work highlighted the significance of mixing conditions for the appearance of state I→state II transition and crazy-clock behavior [13,16]. Therefore, we want to reveal in detail the effect of mixing on the transition, by using stirring bars of different sizes and shapes and applying various mixing rates (Figure 3).

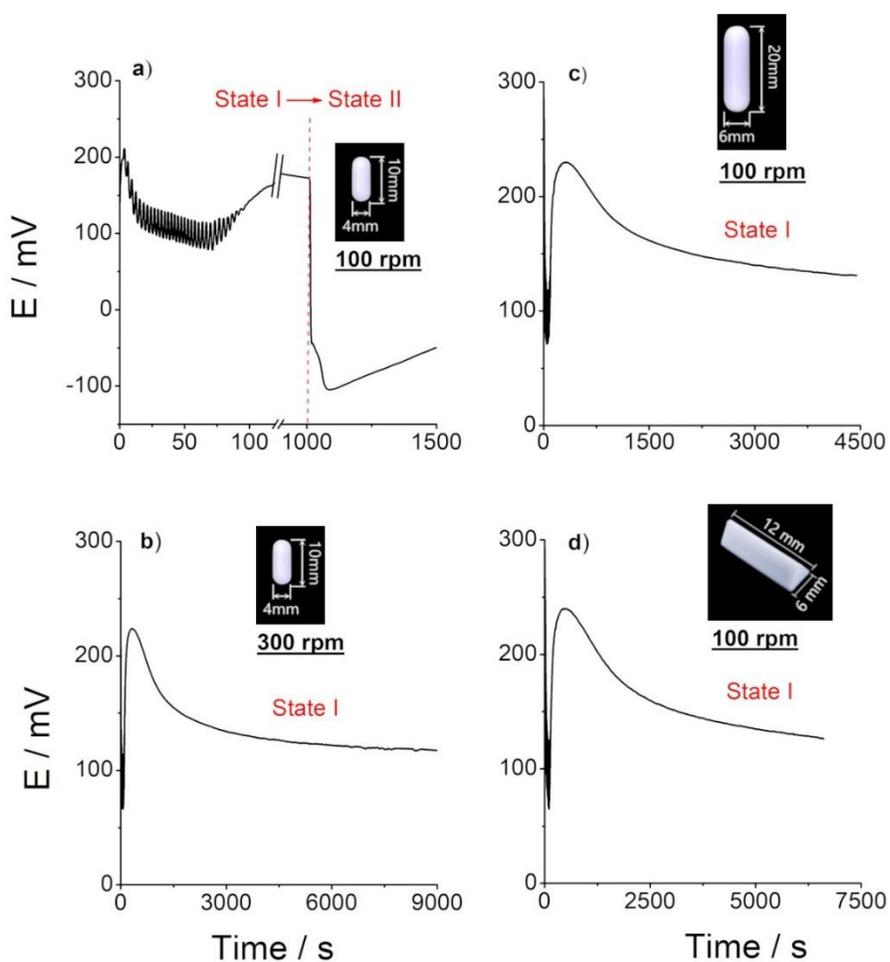

**Figure 3.** Typical measurements with iodide-sensing and reference electrodes with different mixing rates and different shapes of magnetic bar: a) 100 rpm with magnetic stirring bar, PTFE-coated cylindrical, 10 mm length, 4 mm diameter (in inset), b) 300 rpm with magnetic stirring bar, PTFE-coated cylindrical, 10 mm length, 4 mm diameter (in inset), c) 100 rpm with magnetic stirring bar, PTFE-coated cylindrical, 20 mm length, 6 mm diameter (in inset), and d) 100 rpm with magnetic stirring bar, PTFE-coated triangular, 12 mm length, 6 mm diameter (in inset). The reactant concentrations are identical as in Figure 1.

The transition from the state I to state II occurs only with a low stirring rate and a stirring bar of small dimensions (namely, 100 rpm and a magnetic stirring bar made of PTFE-coated cylindrical with a length of 10 mm and a diameter of 4 mm, Figure 3a). The results also underline the importance of the particular magnetic bar shape and mixing rate that was used (Figure 3 a-d). Even with a low stirring rate (100 rpm), the transition does not occur with a bar exhibiting a triangular section (bar 12 mm length, 6 mm diameter) (Figure 3d). This result implies that the transition is strongly connected a particular diffusion conditions and vortex type behavior created by using specific stirring rods. Furthermore, we perform a detailed statistical analysis to reveal the connection between the state I→ state II transition and the mixing rate (using the 10-mm long and 4-mm large cylindrical stirring bar). A set of 60 experiments were performed without stirring and with stirring at a 100-rpm mixing rate (30 experiments, each). The results ($\tau_{osc}$ and $\tau^*$) are tabulated (Table S1 and Table S2) and presented in Supplementary Materials. The $\tau^*$ mean value with 95% confidence limit is for no-mixing $\tau^*_{no\ mix}= (12 \pm 4)$ min and for mixing conditions, $\tau^*_{mix} = (17 \pm 5)$ min.

*3.2. Statistical Analysis of Experimental Results and Evidence of Clustering*

Is there a connection between the time that the system spends in the oscillatory regime and the time when the state I to state II transition occurs? Or, in other words, are the minor differences in oscillatory period duration responsible for a significant deviation in transition appearance? The detailed exploration of the relation between BR oscillatory time, $\tau_{osc}$, and the time $\tau^*$ of the occurrence of the state I→ state II transition (Figure 1), with and without stirring of the solutions, was performed by statistical cluster analysis (CA). Cluster analysis performs subdivision of datasets based on the relationships among their members (in our case datasets of $\tau_{osc}$ and $\tau^*$). The application of CA allows the separation of data in clusters (namely, in groups) based on mutual distances, which reflect a degree of similarity among data [25]. The greater the similarity in the cluster, the higher the distance between the clusters, and hence the better the clustering. Our results combine a total of 60 experiments, obtained with no-mixing conditions (30 experiments) and with a mixing rate of 100 rpm (30 experiments). They were analyzed in the open-source statistic software "R" using "hclust" algorithm for the hierarchical cluster analysis (HCA) [22,23]. The HCA divided the ratios $\tau_{osc}/\tau^*$ into three clusters for both mixing and no-mixing measurements (Figures 4 a and b). It can be noticed that the number of members of a particular cluster slightly changes upon alteration of experimental conditions (Table 1.).

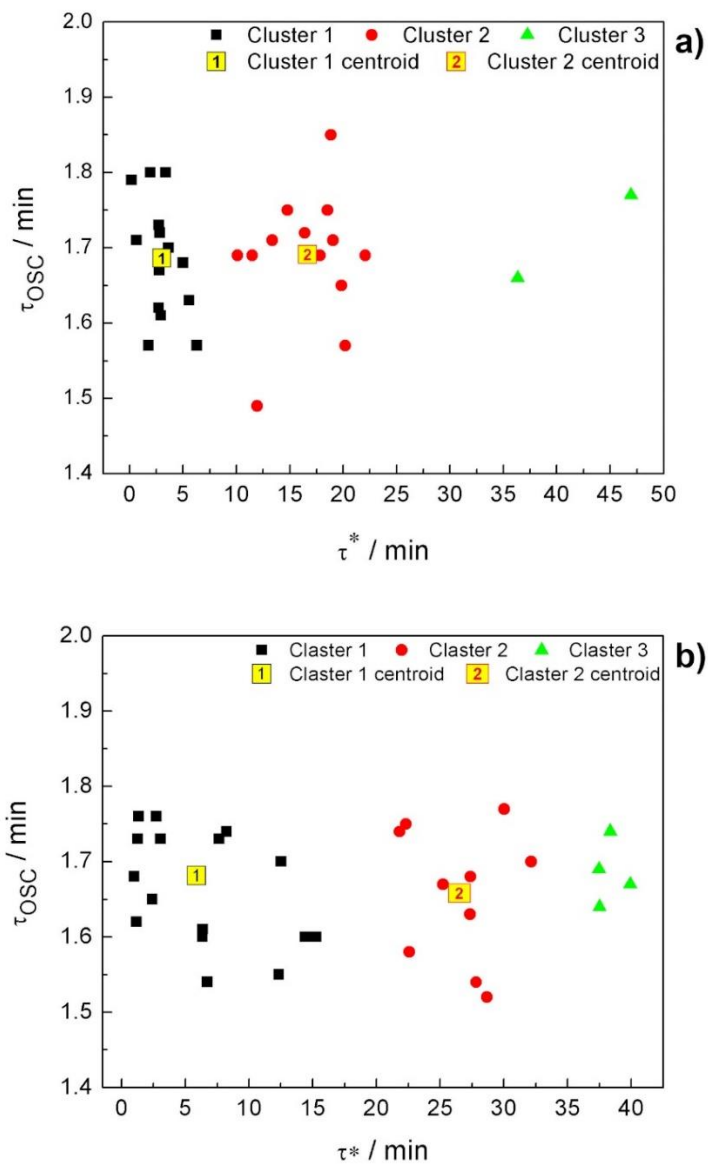

**Figure 4.** Clusters obtained for thirty experiments run without stirring of the reaction mixture a) and clusters obtained for thirty experiments run with stirring of reaction mixture in the vessel, stirring rate was 100 rpm b).

It appears that the stirring effect causes an increase in the members of Cluster 1 and Cluster 3, as well as decreasing in members in Cluster 2. The increase of members in Cluster 3 suggested that stirring has prolonged time for the state I→ state II transition taking place (Figure 4b.). It can be also seen from the $\tau^*$ mean value for no-mixing $\tau^*_{no\ mix}$ = (12 ± 4) min and mixing conditions, $\tau^*_{mix}$ = (17 ± 5) min. Furthermore, all clusters are more compact and separated, in the case of stirring (if compared with those obtained without stirring).

**Table 1.** Clusters and Cluster centroids

| Exp. Condition | Cluster | Number of cluster members | Cluster centroids in minute | |
|---|---|---|---|---|
| | | | $\tau_{osc}$ | $\tau^*$ |
| without stirring | 1 | 15 | 1.686 | 3.036 |
| | 2 | 13 | 1.691 | 16.672 |
| | 3 | 2 | / | / |
| with stirring | 1 | 16 | 1.681 | 5.893 |
| | 2 | 10 | 1.658 | 26.536 |
| | 3 | 4 | / | / |

The results presented in Table 1 clearly indicate that the investigated crazy-clock exhibits a truly random behavior. The shift (Figure 4 a and 4 b, Table 1.) in the position of the cluster centroids towards higher τ* values (i.e., a delaying time to transition to happen) can be observed for the case of applied stirring conditions. Due to a small number of members, the centroid of Cluster 3, in both cases, was not calculated. Furthermore, there is no significant change in the cluster's centroid position regarding $\tau_{osc}$ oscillatory coordinate for clusters. That leads to the conclusion that the transition from state I to state II is independent of oscillatory time duration or, in other words, that the minor differences in oscillatory period duration are not responsible for a significant deviation in transition appearance. Therefore, we calculated the optimal number of clusters by a one parameter (by parameter τ*) K-means analysis, for no-mixing and mixing conditions (Figure 5 a and b, respectively). The determination of optimal numbers of clusters was performed using gap method in fviz_nbclust alghoritm [26]. In brief, the gap statistic compares the total within intra-cluster variation for different values of k with their expected values under null reference distribution of the data. The estimate of the optimal clusters will be value that maximizes the gap statistic (i.e that yields the largest gap statistic) [27].

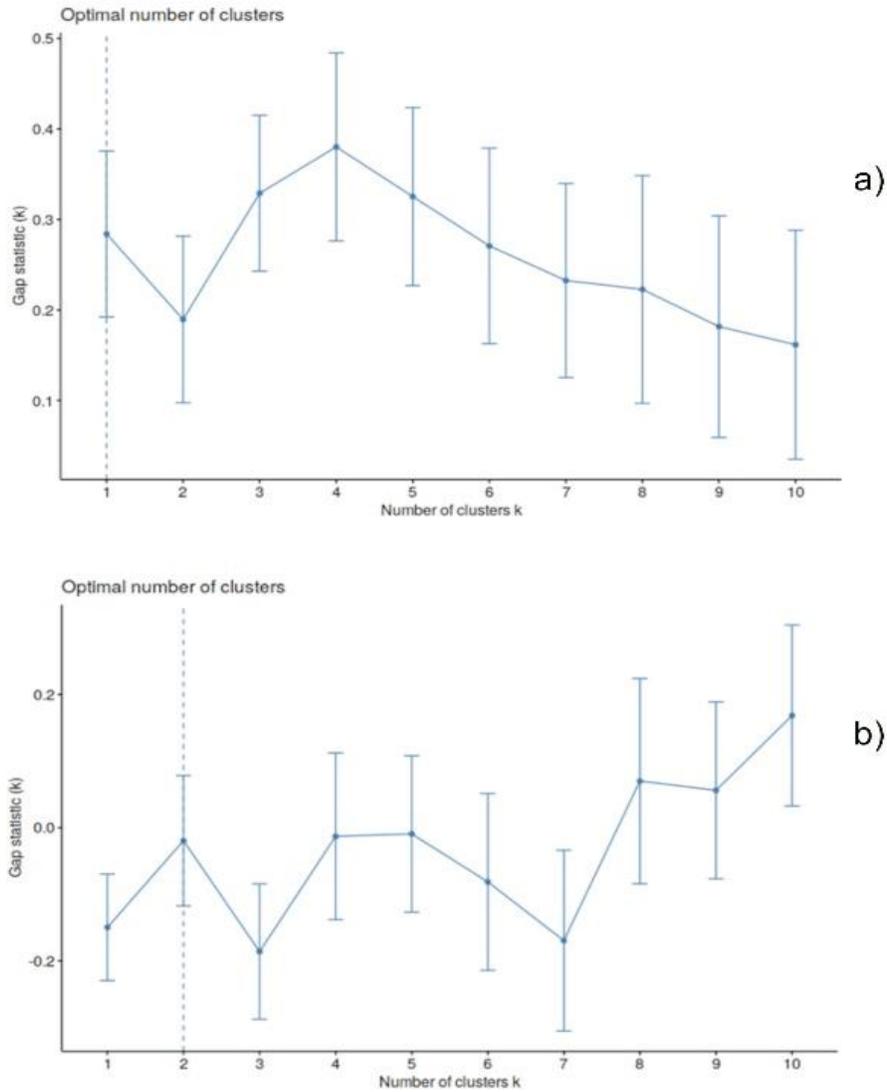

**Figure 5.** The optimal number of clusters for one parameter analysis by parameter $\tau^*$ (time when state I to state II transition occurs) for non-mixing conditions a) and mixing conditions 100 rpm b).

As it can be seen from the Figure 5 the optimal number of clusters obtained by using one parameter K-means analysis is changed. The BR system which is not stirred has one cluster (Figure 5a), while the stirring induces differentiation of two clusters (see Figures 5b and 6). The clusters are well separated. This indicates that mixing introduced additional effects responsible for a significant cluster separation revealing the existence of time domains where the state I to state II transition is more likely to occur.

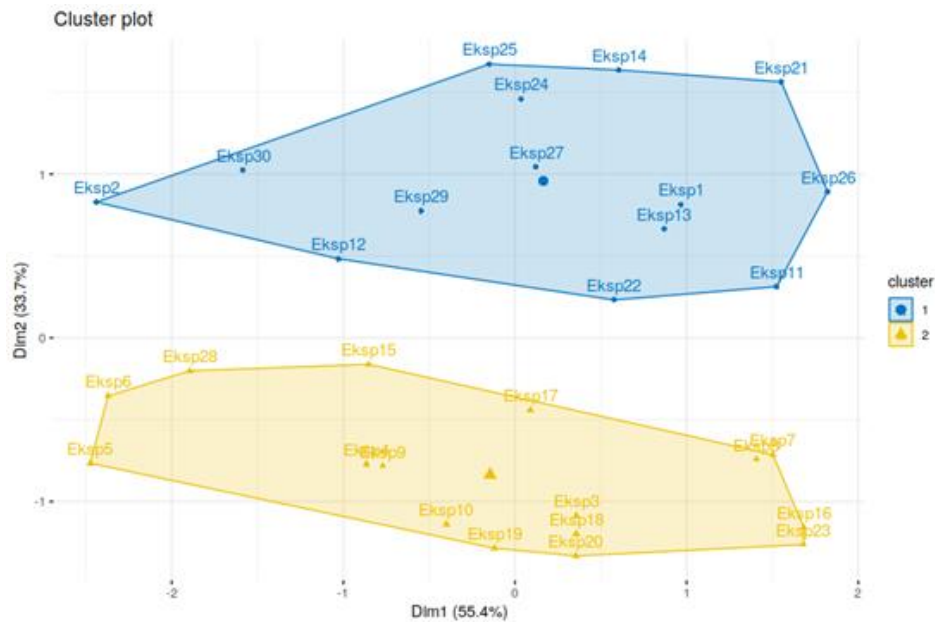

**Figure 6.** The evidence of two clusters for state I→state II transition obtained for mixing condition in BR system. Clusters are plotted in two dimensions (Dim1 and Dim2).

*3.3. The state I → state II phenomenon and its relation to (spontaneous) symmetry breaking*

The appearance of clusters indicates the existence of time domains where the state I to state II transition is more likely to occur. The stirring of the reaction mixture has a strong indirect influence on the state I→state II transition, delaying the crazy-clock behavior, shifting cluster centroids toward higher τ* values, and increasing the cluster separation (i. e. time domain separation), as well. The existence of clusters could be connected to different nucleation and growth mechanisms of iodine crystals in the case of mixing [28], and further examination of solid iodine products would be the subject of future work.

However, if we assumed that state I (low iodide and iodine concentration) is the symmetric state, which under some conditions becomes absolutely unstable, then, reaching the state II (high iodide and iodine concentration, with a new $I_2$ solid phase) could be considered as spontaneous symmetry breaking [29], see Figure 7. Such an observation of state I to state II transition could also explain the persistence of the BR system "indefinitely" in the state I, as obtained for strong mixing condition (see Figure 3).

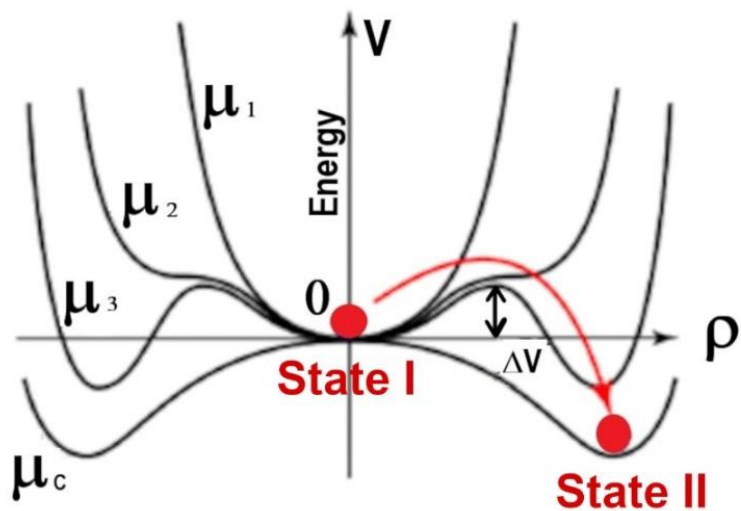

**Figure 7.** Posible symmetry breaking during the transition from the state I (low iodide and iodine concentration) to state II (high iodide and iodine, with segregation of solid iodine) in Briggs-Rauscher oscillatory reaction after an oscillatory period. Energetic consideration and symmetry breaking depending on control parameter, µ. ΔV corresponds to the energy threshold of the formation of solid iodine from chemical reactions that occur in the BR solution after oscillatory period.

Namely, in the case of the symmetry-breaking process, the system must overcome a sufficiently large energy barrier (as shown in Figure 7). Therefore, the system's state (state I, low iodide and iodine concentration) will remain unchanged until a sufficiently large perturbation throws it over the energy barrier, ΔV, which separates the states. We assume that ΔV corresponds to the energy threshold of the formation of solid iodine from chemical reactions that principally occur in the BR solution after the oscillatory period. It is well-known that the post-oscillatory period could be excitable [30,31], meaning that a nonlinear system can be shifted (perturbed) from one state to another. It is usually achieved by the addition of some stable intermediate or reactants, playing the role of an external perturbation, to the reaction mixture [32]. Since, in our case, there is no external perturbants/ stimulus (all experiments are conducted under identical conditions, and the system remained under constant temperature), the BR system should find an internal stimulus to overcome the energy barrier. The mixing itself should not influence activation energies of chemical reactions responsible for state I to state II transition. In other words, the energy threshold ($\Delta V$) should be identical for mixing and no-mixing condition. However, the system has behaved differently, and the state I to state II transition strongly depends on the mixing conditions. Therefore, some phenomena related to mixing are accountable for the obtained behavior.

The diffusion-driven instability combined with nonlinear chemical reactions (with autocatalytic steps and radical reactions) is a broad concept and it could be responsible for the "internal stimulus" necessary for passing the energy barrier. The diffusion-driven instability is intensified by gaseous oxygen and carbon dioxide/carbon monoxide, which are released in the BR solution during the oscillatory period [33]. Additionally, the possible energetic coupling between physical and chemical processes, such as the nucleation of gaseous phase ($O_2$ and/or $CO_2$), nucleation of solid iodine and particular chemical reactions, could also be considered as an "internal stimulus" necessary for overcoming the energy barrier and breaking symmetry [34-36].

This proposal is actually a reformulated original idea of Turing [37], where the interplay of chemical reactions and diffusion are responsible for pattern formation in living systems. As suggested by Prigogine, the spontaneous appearance of a spatial organization via diffusion-driven instability can be considered as a spontaneous symmetry-breaking transition [38]. In the presented work, we deal with a bulk solution and there is no visible spatial organization, but the spatial organization (spatiotemporal patterns) of the same process (state I→state II transition in Briggs-Rauscher reaction) in a thin layer, is very recently found by Epstein and coworkers [15]. Therefore, this work indirectly links spontaneous symmetry breaking and crazy-clock behavior (stochastic nature) in the bulk. The stochastic nature of state I to state II transition and its relation to symmetry breaking (and pattern formation), introduced a new approach in the investigation of crazy-clock behavior. On the other hand, the investigation of chemical systems with stochastic nature and symmetry breaking could improve our understanding of more complex phenomena in living organisms, such as morphogenesis.

Additionally, this paper nominates state I→state II transition as an easily available chemical system for intrinsic random number generator and thus, expands the potential application of this crazy-clock reaction.

4. **Conclusions**

In this work, we further investigated the crazy-clock phenomenon (state I to state II transition) which occurs after a strongly reproducible Briggs-Rauscher oscillatory reaction. The mixing rate, as well as the magnetic bar shape and dimensions, have a strong influence on the transition appearance. In order to better understand the stochasticity of the mentioned process, we ran more than 60 experiments (30 experiments with mixing and 30 experiments with no-mixing conditions), and we applied the statistical cluster K-means analysis. Although the transition for both mixing and no-mixing conditions are taking place completely randomly, by using statistical cluster analysis, we obtained different number of clusters pointing to different time-domains where the transition is more likely to occur. Two-parameter analysis (by oscillatory time duration, $\tau_{osc}$ and by the moment when the transition occurs, $\tau^*$) suggests that the state I→state II is independent of the oscillatory time duration. Therefore, we performed a one parameter analysis (by $\tau^*$). In the case of no-mixing, we found one cluster, while the statistical analysis of the results for mixing conditions revealed two compact and well-separated clusters. The clustering method reveals new hidden details regarding the chemical dynamics of nonlinear processes. The state I to state II transition could be explained through a symmetry breaking approach, and the necessity of the BR system to overcome a sufficiently large energy barrier. This is the first link of the crazy-clock behavior to the symmetry breaking phenomenon. The investigation of chemical systems with stochastic nature and symmetry breaking could improve our understanding of more complex phenomena in living organisms and therefore the scope of the presented results goes beyond oscillatory reaction kinetics. Furthermore, the described example belongs to the small class of chemical systems that shows intrinsic randomness in their response and it might be considered as a real example of a classical liquid random number generator.

**Supplementary Materials:** A set of 60 experiments were performed without stirring and with stirring at a 100-rpm mixing rate (30 experiments, each). The results ($\tau_{osc}$, number of oscillations and τ*) are tabulated (Table S1 and Table S2) and presented in Supplementary Materials. The supporting information can be downloaded at: www.mdpi.com/xxx/s1.

**Author Contributions:** Conceptualization, M.P. and B.K.; methodology, M.P., J.M., Y.C., and B.K.; statistical K-means analysis M.D.; investigation, M.P. and J.M.; writing—original draft preparation, M.P. and B.K; writing—review and editing, M.P., B.K., Y.C., S.M. and T.V.; visualization, J.M. and B.B..; All authors have read and agreed to the published version of the manuscript.

**Data Availability Statement:** Data underlying the results presented in this paper are not publicly available at this time but may be obtained from the authors upon reasonable request. Please contact for this corresponding authors.

**Acknowledgments:** This work was supported by the Ministry of Education, Science and Technological Development of the Republic of Serbia Contract numbers: 451-03-9/2021-14/200026 and 451-03-9/2021-14/200146. SRM was supported by a BEWARE Fellowship of the Walloon Region (Convention n°2110034), as a postdoctoral researcher. Y.C. is a research associate of the Fund for Scientific Research F.R.S.-FNRS. B. K. and B. B. acknowledge financial support of the Ministry of Science, Republic of Serbia (grant III 45016). Additionally B. K. acknowledges support from F R S – FNRS.

**Conflicts of Interest:** The authors declare no conflict of interest.